\documentclass[preprint,12pt]{elsarticle}

\usepackage{amssymb}
\usepackage[]{amsmath}
\usepackage{mathrsfs}
\usepackage{mathtools} 
\usepackage{enumerate}
\usepackage{caption}
\usepackage[dvipsnames,svgnames,table]{xcolor}
\usepackage{IEEEtrantools}
\usepackage{graphicx}% Include figure files
\usepackage{subcaption}
\usepackage{ragged2e} % for justifying
\usepackage{array}
\usepackage{multirow}
\usepackage{verbatim}
\usepackage{soul}
\usepackage{tabularx}
\usepackage{booktabs,siunitx}
\usepackage[T1]{fontenc}
\usepackage{hyperref}
\usepackage{physics}
\usepackage{siunitx}
\usepackage{makecell}
\usepackage{float}
\usepackage{adjustbox}

\journal{Reliability Engineering \& System Safety}

\begin{document}

\begin{frontmatter}

\title{Topology and Fragility of European High-Voltage Networks: A Cross-Country Comparative Analysis} %% Article title

\author[label1]{Bálint Hartmann}
\affiliation[label1]{organization={Department of Electric Power Engineering, Budapest University of Technology and Economics,},
            addressline={Egry J. u. 18.},
            city={Budapest},
            postcode={H-1111},
            country={Hungary}}
\author[label2,label3]{Michelle T. Cirunay}
\affiliation[label2]{organization={Institute of Technical Physics and Materials Science, HUN-REN Centre for Energy Research},
            addressline={P.O. Box 49},
            city={Budapest},
            postcode={H-1525},
            country={Hungary}}
\affiliation[label3]{organization={Dr. Andrew L. Tan Data Science Institute, De La Salle University},
            addressline={Taft Avenue},
            city={Manila},
            postcode={2401},
            country={Philippines}}

%% Abstract
\begin{abstract}
Reliable electricity supply depends on the seamless operation of high-voltage grid infrastructure spanning both transmission and sub-transmission levels. Beneath this apparent uniformity lies a striking structural diversity, which leaves a clear imprint on system vulnerability. In this paper, we present harmonized topological models of the high-voltage grids of 15 European countries, integrating all elements at voltage levels above 110 kV.

Topological analysis of these networks reveals a simple yet robust pattern: node degree distributions consistently follow an exponential decay, but the rate of decay varies significantly across countries. Through a detailed and systematic evaluation of network tolerance to node and edge removals, we show that the decay rate delineates the boundary between systems that are more resilient to failures and those that are prone to large-scale disruptions. Furthermore, we demonstrate that this numerical boundary is highly sensitive to which layers of the infrastructure are included in the models.

To our knowledge, this study provides the first quantitative cross-country comparison of 15 European high-voltage networks, linking topological properties with vulnerability characteristics.
\end{abstract}

%%Graphical abstract
\begin{graphicalabstract}
\begin{figure}[H]
     \centering    \includegraphics[width=\columnwidth ]{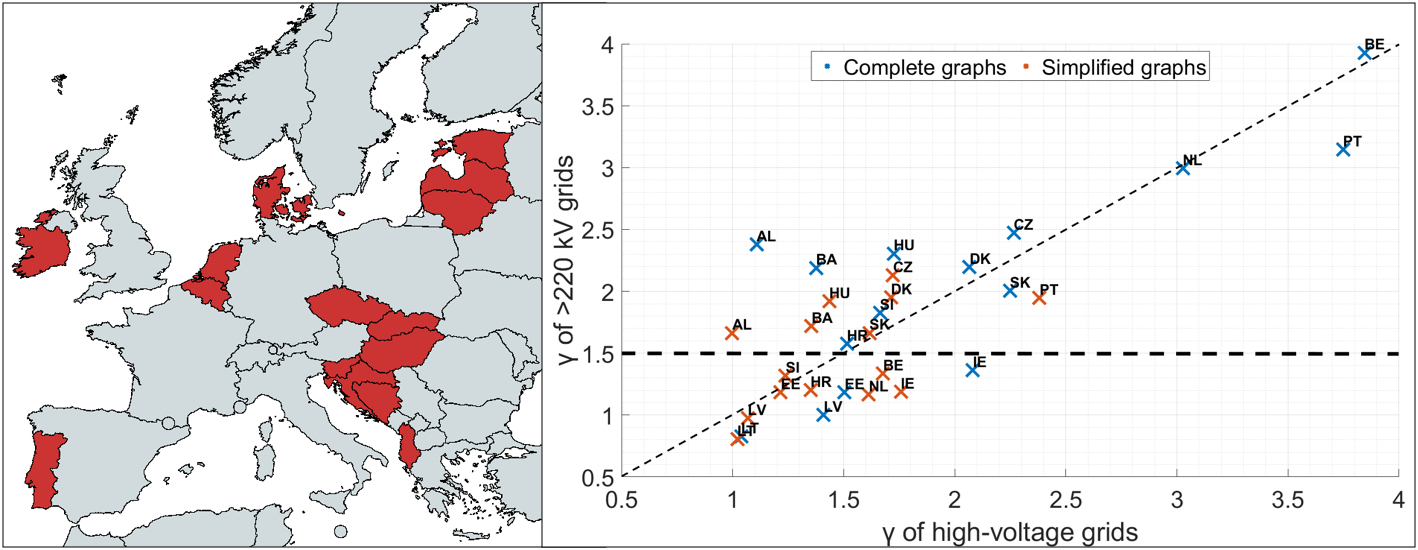}
     %\caption{}
     \label{fig:Graph_abstract}
 \end{figure}
\end{graphicalabstract}

%%Research highlights
\begin{highlights}
\item Harmonized models of 15 European high-voltage grids above 110 kV
\item Node degree distributions follow exponential decay across all countries
\item Decay rate separates resilient and fragile systems under failures
\item Sub-transmission layers significantly affect topological metrics
\end{highlights}

%% Keywords
\begin{keyword}
Power grid topology \sep High-voltage networks \sep Node degree distribution \sep Network vulnerability \sep Performance
\end{keyword}

\end{frontmatter}

\section{Introduction}
\label{sec:sec1}
In 1999, Barabási and Albert reported that large complex networks show a high level of self-organization, coming from the way nodes interact. This property became known as scale-free behavior \cite{Barabasi1999}. They showed their ideas on real-world networks, often using the electrical power grid of the Western United States (where nodes are generators, transformers, and substations, and edges are the power lines connecting them). Their work received a lot of attention from the scientific community and led to many studies that questioned their first results, especially about whether the node degree distribution really follows a power-law. Amaral et al. pointed out that in power grids the degree distribution matches better with an exponential distribution than with a power-law, especially for nodes with small degrees \cite{Amaral2000}. They suggested that aging and limited node capacity could be the reason for this difference.
In the first decade of the new millennium, many papers were written on this topic. Researchers studied small-world behavior in different power grids, including the Western US \cite{Holmgren2006,ZhifangWang2010}, North America \cite{KeSun2005,Kim2007}, China \cite{KeSun2005,Ding2006,Mei2011}, Scandinavia \cite{Holmgren2006}, Europe \cite{ROSASCASALS2007}, and The Netherlands \cite{Pagani2011}. They fitted different distributions to the cumulative probability of node degrees, such as exponential \cite{KeSun2005,ROSASCASALS2007,Crucitti2004,Sol2008,RosasCasals2009,Espejo2018,Hartmann2021}, power-law \cite{Chassin2005}, and mixed \cite{ZhifangWang2010,Pagani2011,Rosato2007,Wang2010} models. As of now, exponential fits are the most widely accepted models and have become a core insight in the complex network modeling of power grids.

Using a common distribution form it was the paper of Solé et al. \cite{Sol2008}, which compared countries' grid structures to identify patterns and vulnerabilities using percolation theory. They analyzed the UCTE European grid dataset (230 nodes, 400 links). Plotting the degree distribution $P(k)=C \cdot e^{-k/\gamma} $, they introduced a classification of European grids into "robust" and "fragile" categories based on the exponential decay constant $\gamma$. They found a critical threshold around $\gamma \approx 1.5$: grids with $\gamma <1.5$ (e.g. Belgium, the Netherlands, Germany, Italy) showed better robustness, while those with $\gamma >1.5$ (e.g. France, Hungary, Spain, Serbia) were more vulnerable. This grouping of topologies correlated with real-world reliability statistics of the corresponding power systems.

In a related work, Rosas-Casals et al. \cite{ROSASCASALS2007} also formed two clusters of European power grids based on their vulnerability and showed that certain topological metrics (e.g. average degree, redundancy of connectivity) align with empirical reliability indicators. They extended their work with motif analysis these networks \cite{RosasCasals2009}, observing that more fragile networks (ones with higher $\gamma$) included a larger share of high-connectivity motifs (stars, triangles) compared to robust networks. They concluded that highly connected motifs may improve reliability locally but also concentrate critical hubs.

These comparative studies demonstrated that while European power grids show similar characteristics, their topology differs enough to be grouped into classes with different vulnerability levels, and the classification threshold of $\gamma = 1.5$ has become a standard of the field. However, mostly due to the lack of data, these studies only evaluated transmission grids over 220 kV, which in general have better reliability than the complete high-voltage network including transmission and sub-transmission levels \cite{Hadaj2020,ENTSOE2024}.

The aim of present paper is to close this gap by presenting a cross-system analysis of 15 European high-voltage networks. The dateset compiled to this study includes sub-transmission voltage levels, i) containing an order of magnitude more nodes and edges, ii) introducing radial and semi-radial structures and iii) largely changing local connectivity patterns. These three aspects significantly affect the node degree distribution and lead to different $\gamma$ values. Sub-transmission layers of power grids are responsible for handling regional redundancies and the isolation of faults. However, countries in Europe have different philosophies for planning these networks, which can also be revealed through the analysis of $\gamma$ values.

The remainder of the paper is organized as follows. Section \ref{sec:sec2} introduces the topological dataset, the metrics used for comparison and the methods for determining the performance of the networks. These results are presented in \ref{sec:sec3}, and discussed in \ref{sec:sec4}. Finally, conclusions are drawn in \ref{sec:sec5}.

\section{Scope and Data}
\label{sec:sec2}

In this section, we present the datasets being considered and the corresponding links to their repositories, and the metrics by which their performances are measured.

\subsection{Topological data}\label{sec2.1}
The present paper analyzes topological data of 15 European countries: Albania (AL), Bosnia and Herzegovina	(BA), Belgium (BE), Czechia (CZ), Denmark (DK), Estonia (EE), Croatia (HR), Hungary (HU), Ireland (IE), Lithuania (LT), Latvia (LV), the Netherlands (NL), Portugal (PT), Slovenia (SI) and Slovakia (SK). The datasets were compiled by the first author from heterogeneous sources, including statistical yearbooks, cartographic materials, and official data publications. A rigorous pre-processing and standardization was carried out to ensure cross‑source. The dataset is openly available through the ARP Research Data Repository \cite{handleTransmissionSubtransmission}.

To ensure comparability with studies published in the literature, different modeling complexities were used, when calculating $\gamma$. The default network representations contain all voltage levels (sub-transmission and transmission) and all elements (i.e. redundant lines are not simplified). The default representations were modified in two ways. In one modification, the sub-transmission (<220 kV) elements were removed from the representation. in another modification, the graph representations were simplified so that redundancies are removed.

Doing so, the network representations after the second modification have the same modeling complexity as the ones discussed by Solé et al. \cite{Sol2008} and Rosas-Casals et al. \cite{ROSASCASALS2007}, while the default networks are an accurate representation of the actual grids. Another reason for calculating the $\gamma$ values is to provide insight to how the addition of lines and/or multiple voltage levels affect the node degree distributions and thus the vulnerability of power grids.
In general, increasing $\gamma$ values show a less steeper distribution with a flattening tail, which is expected from the inclusion of sub-transmission topologies. On the other hand, the simplification of the networks through removing the parallel circuits usually only decreases the degrees of large hubs, decreasing $\gamma$. In the literature this shift is reported to be between 5-15$\%$, but the effect largely depends on the voltage layer being studied \cite{Rosato2007,Bompard2011,Schultz2014,Panigrahi2020}. Sub-transmission networks are more meshed, and thus have a higher edge density.

\subsection{Topological metrics}\label{sec2.2}

In this work, all the considered datasets are undirected and unweighted networks. To provide a hint on their structure, we utilized various topological metrics. In following, the mathematical descriptions for each and their implications are presented.

The link density $D$ measures the ratio of existing edges relative to the maximum possible number of edges in a fully connected network:

\begin{equation}
D = \frac{2E}{N(N-1)}
\label{eqn:density}
\end{equation}

\noindent where $N$ is the number of nodes and $E$ is the number of edges.

The average node degree for an undirected graph is simply the average number of edges per node in the graph:

\begin{equation}
\langle k \rangle = \frac{2E}{N}
\label{eqn:ave_degree}
\end{equation}

\noindent where the factor of 2 arises from each edge contributing to the degree of two distinct vertices.

The average path length $L$ is defined as the average number of steps along the shortest paths between all pairs of nodes:

\begin{equation}
L = \frac{1}{N(N-1)} \sum_{i \neq j} d(i,j)
\label{eqn:spl}
\end{equation}

\noindent where $d(i,j)$ is the shortest path between nodes $i$ and $j$.

Another metric to characterize the network is its diameter $d$ which is the longest among all calculated shortest paths in the network. As this quantifies the distance between the two most distant nodes in the network, the diameter of the graph provides a hint on the overall connectivity i.e. a smaller diameter implies better connectivity across the network.

The clustering coefficient $C$ measures the tendency of a node’s neighbors to be connected and is given by:

\begin{equation}
C = \frac{1}{N} \sum_{i} \frac{2E_i}{k_i(k_i - 1)}
\label{eqn:clustering}
\end{equation}

\noindent where $E_i$ is the number of edges between the neighbors of node $i$, and $k_i$ is its degree.

To detect sections of the networks that are highly connected locally but have sparse connections to other clusters, we employ the modularity metric $Q$ \cite{Newman2006-bw} given by

\begin{equation}
Q = \frac{1}{2E} \sum_{ij} \left( A_{ij} - \gamma \frac{k_i k_j}{2E} \right) \delta(c_i,c_j)
\label{eqn:modularity}
\end{equation}

\noindent where $E$ is the total number of edges in the network, $A_{ij}$ is an element of the adjacency matrix (1 if there is an edge between nodes $i$ and $j$), $k_i$ is the degree of node $i$, $c_i$ is the community where node $i$ belongs to, and $\delta(c_i, c_j)$ is the Kronecker delta function whose value is 1 if nodes $i$ and $j$ belong to the same community, and 0 otherwise. If $Q > 0$ indicates good community structure; $Q \approx 0$ no community structure is detected; and if $Q < 0$, although rare, indicates the network is less modular than random.

To provide a measure of performance of the networks as nodes and edges are randomly removed, we consider the global efficiency $eff$ which is a measure of efficient power flows in the network \cite{latora2001}. Importantly, efficiency is a robust metric, as its validity is not depending on network size or other contextual factors. It is defined as:

\begin{equation}
eff = \frac{1}{N(N-1)} \sum_{i \neq j} \frac{1}{d(i,j)}
\label{eqn:efficiency}
\end{equation}

\noindent High $eff$ means that most nodes can reach others easily, even in large networks.

The Watts-Strogatz small-world coefficient $\sigma$, proposed by Fronczak et al. \cite{FronczakPRE2004}, combines clustering and path length to characterize small-world behavior,

\begin{equation}
\sigma = \frac{C/C_r}{L/L_r}
\label{eqn:sigma}
\end{equation}

\noindent where the $C_r$ and $L_r$ are the corresponding clustering coefficient and average shortest path length for random graphs \cite{FronczakPRE2004}, given by:

\begin{equation}
C_r = \frac{\left< k \right>}{N}
\label{eqn:clustering_random}
\end{equation}

\begin{equation}
L_r = \frac{\ln(N) - 0.5772}{\ln \left< k \right>} + 0.5
\label{eqn:spl_random}
\end{equation}

\noindent A network is considered small-world if $\sigma > 1$, meaning it has significantly higher clustering than a random network while maintaining a comparable path length.

An alternative small-world measure, proposed by Telesford et al. \cite{TelesfordBRAIN2011}, addresses the limitations of $\sigma$ by comparing the clustering coefficient to that of a lattice (regular) network. This metric, $\omega$, ranges between -1 and 1:

\begin{equation}
\omega = \frac{L_r}{L} - \frac{C}{C_{\text{lattice}}}
\label{eqn:omega}
\end{equation}

\noindent Values of $\omega \approx 0$ indicate small-world characteristics ($L \approx L_r$, $C  \approx C_{lattice}$). When $\omega > 0$ the network is more random-like ($L \approx L_r$, $C <  C_{lattice}$), while $\omega < 0$ indicates a more regular, lattice-like topology ($L > L_r$, $C  \approx C_{lattice}$)

\subsection{Performance}\label{sec2.3}
Performance of the 15 networks was evaluated using four performance metrics, while certain percentage of nodes or edges were randomly removed from the grid. The four performance metrics are:
\begin{itemize}
\item Share of edges lost: as nodes or edges are randomly removed from the network, additional edges may also lose their connectivity. The larger this share, the more vulnerable is the grid.

\item Largest connected component (LCC): the set of nodes within a graph that has the greatest number of nodes, where every node can be reached from every other node by following the graph's edges. In undirected graphs, like in this study, this is found by identifying all separate connected groups of nodes and selecting the one with the most nodes. The bigger is the largest connected component after the removal, the more intact is the structure of the network.

\item The change of efficiency ($\Delta eff_-/eff_0$): as efficiency is a measure of the network's performance, thus the smaller is the specific decrease of efficiency (determined as the ratio of efficiency decrease and initial efficiency), the more tolerant is the network to removals.

\item The change of clustering($\Delta C/C_0$): randomly removed nodes or edges can significantly affect the underlying structure of the network. This is measured by the specific decrease of the clustering coefficient. Similarly to the change of efficiency, smaller values indicate better performance.

\end{itemize}

Values of the four performance metrics were calculated and standardized for a total of 10 scenarios (5 node removals, 5 edge removals). The composite performance of the 15 networks was calculated as the sum of these standardized values, where larger numbers are connected to better overall performance.

For each examined network, 10 scenarios were defined, in which 1, 2, 5, 10, and 20$\%$ of nodes or edges were removed from the network, respectively. The removed elements were selected randomly to create an even mapping of the event horizon. The number of runs was selected aiming for numerical convergence of the resulting histograms, thus providing good statistical representation; each scenario was run 10,000 times. All calculations were carried out using MATLAB R2022b.

\section{Results}
\label{sec:sec3}
In the following, results of the analysis are summarized along three aspects: topological metrics, the effect of node/edge removals, and 
performance.

\subsection{Topological metrics}\label{sec3.1}

Table~\ref{tab:network-metrics} summarizes the most important metrics and topological characteristics of the 15 networks.

\begin{table}[ht]
  \centering
  \caption{Network metrics by country} \label{tab:network-metrics}
  \resizebox{\textwidth}{!}{%
  \begin{tabular}{|r|rrrrrrrrrrrrrrr|}
    \toprule
    & \textbf{AL} & \textbf{BA} & \textbf{BE} & \textbf{CZ} & \textbf{DK} & \textbf{EE} & \textbf{HR} & \textbf{HU} & \textbf{IE} & \textbf{LT} & \textbf{LV} & \textbf{NL} & \textbf{PT} & \textbf{SI} & \textbf{SK} \\
    \midrule
    $N$ & 133 & 188 & 281 & 521 & 204 & 153 & 241 & 387 & 306 & 265 & 147 & 311 & 145 & 136 & 249 \\
    $E$ & 175 & 260 & 666 & 1019 & 307 & 231 & 358 & 640 & 467 & 329 & 220 & 669 & 314 & 233 & 457 \\
    $D$ & 0.020 & 0.015 & 0.017 & 0.008 & 0.015 & 0.020 & 0.012 & 0.009 & 0.010 & 0.009 & 0.021 & 0.014 & 0.030 & 0.025 & 0.015 \\
    $\langle k \rangle$ & 2.632 & 2.766 & 4.740 & 3.912 & 3.010 & 3.020 & 2.971 & 3.307 & 3.052 & 2.483 & 2.993 & 4.302 & 4.331 & 3.426 & 3.671 \\
    $d$ & 23 & 13 & 18 & 13 & 20 & 19 & 29 & 14 & 16 & 22 & 13 & 16 & 15 & 12 & 15 \\
    $L$ & 6.945 & 5.970 & 7.601 & 5.678 & 8.113 & 6.500 & 8.570 & 6.566 & 7.199 & 8.391 & 6.036 & 7.465 & 5.936 & 5.177 & 5.360 \\
    $C$ & 0.035 & 0.066 & 0.151 & 0.172 & 0.107 & 0.090 & 0.085 & 0.077 & 0.099 & 0.050 & 0.082 & 0.173 & 0.270 & 0.237 & 0.227 \\
    $Q$ & 0.433 & 0.451 & 0.469 & 0.424 & 0.496 & 0.345 & 0.230 & 0.467 & 0.466 & 0.457 & 0.415 & 0.460 & 0.469 & 0.461 & 0.471 \\
    $\sigma$ & 1.256 & 3.791 & 4.409 & 18.774 & 4.303 & 3.192 & 4.035 & 6.855 & 7.040 & 3.795 & 3.016 & 6.777 & 5.331 & 7.301 & 12.329 \\
    $\omega$ & 0.591 & 0.664 & 0.035 & 0.423 & 0.263 & 0.442 & 0.314 & 0.521 & 0.419 & 0.523 & 0.542 & 0.087 & -0.118 & 0.252 & 0.296 \\
    $eff$ & 0.195 & 0.202 & 0.164 & 0.198 & 0.172 & 0.199 & 0.164 & 0.178 & 0.170 & 0.151 & 0.208 & 0.164 & 0.219 & 0.236 & 0.221 \\
    Share of 110–150 kV lines & 0.806 & 0.854 & 0.782 & 0.917 & 0.808 & 0.931 & 0.922 & 0.905 & 0.824 & 0.930 & 0.905 & 0.857 & 0.318 & 0.936 & 0.902 \\
    Share of 220–275 kV lines & 0.177 & 0.108 & 0.105 & 0.026 & 0.026 & 0.000 & 0.064 & 0.039 & 0.167 & 0.000 & 0.000 & 0.045 & 0.430 & 0.017 & 0.018 \\
    Share of 330–400 kV lines & 0.017 & 0.038 & 0.113 & 0.058 & 0.166 & 0.069 & 0.014 & 0.056 & 0.009 & 0.070 & 0.095 & 0.099 & 0.252 & 0.047 & 0.081 \\
    $\gamma$ of HV grid exp. fit & 1.108 & 1.376 & 3.845 & 2.266 & 2.066 & 1.503 & 1.514 & 1.726 & 2.081 & 1.039 & 1.409 & 3.027 & 3.748 & 1.663 & 2.248 \\
    $\gamma$ of HV grid simple graph exp. fit & 0.998 & 1.354 & 1.677 & 1.721 & 1.713 & 1.215 & 1.351 & 1.435 & 1.757 & 1.021 & 1.068 & 1.612 & 2.380 & 1.238 & 1.616 \\
    $\gamma$ of >220 kV grid exp. fit & 2.379 & 2.189 & 3.928 & 2.473 & 2.196 & 1.184 & 1.577 & 2.305 & 1.363 & 0.828 & 1.001 & 2.993 & 3.146 & 1.827 & 2.004 \\
    $\gamma$ of >220 kV grid simple graph exp. fit & 1.663 & 1.719 & 1.336 & 2.128 & 1.949 & 1.184 & 1.201 & 1.918 & 1.189 & 0.803 & 0.973 & 1.166 & 1.945 & 1.319 & 1.663 \\
    \bottomrule
  \end{tabular}}
\end{table}

Here we see that the selected European high-voltage grids are sparse networks with node counts ranging from 133 to 521 and edge densities (0.008–0.03), yielding average degrees between 2.5 and 4.7 and moderate path lengths ($L  \sim 5 - 8$) despite relatively long diameters ($d \sim 12 - 29$). The clustering coefficients are low ($C \sim 0.035 - 0.27$), indicating locally tree-like structures, while the modularity ($Q \sim 0.23 - 0.49$) suggests the presence of identifiable communities. Meanwhile, the voltage hierarchy is dominated by 110–150 kV sub-transmission lines, with >220 kV lines forming a smaller backbone that supports robustness, particularly in countries with a larger fraction of 220–400 kV lines. Finally, the Telesford small-worldness ($\omega$ ranging from slightly negative to positive) shows that the networks lie near the small-world regime, indicating moderate global efficiency ($eff \sim 0.15 - 0.24$) and local clustering. The variability in the Watts-Strogatz small-world coefficient $\sigma$ reflects differences in the extent to which each network exhibits small-world characteristics i.e. higher $\sigma$ indicates a stronger small-world structure.

\begin{figure}[H]
     \centering    \includegraphics[width=\columnwidth ]{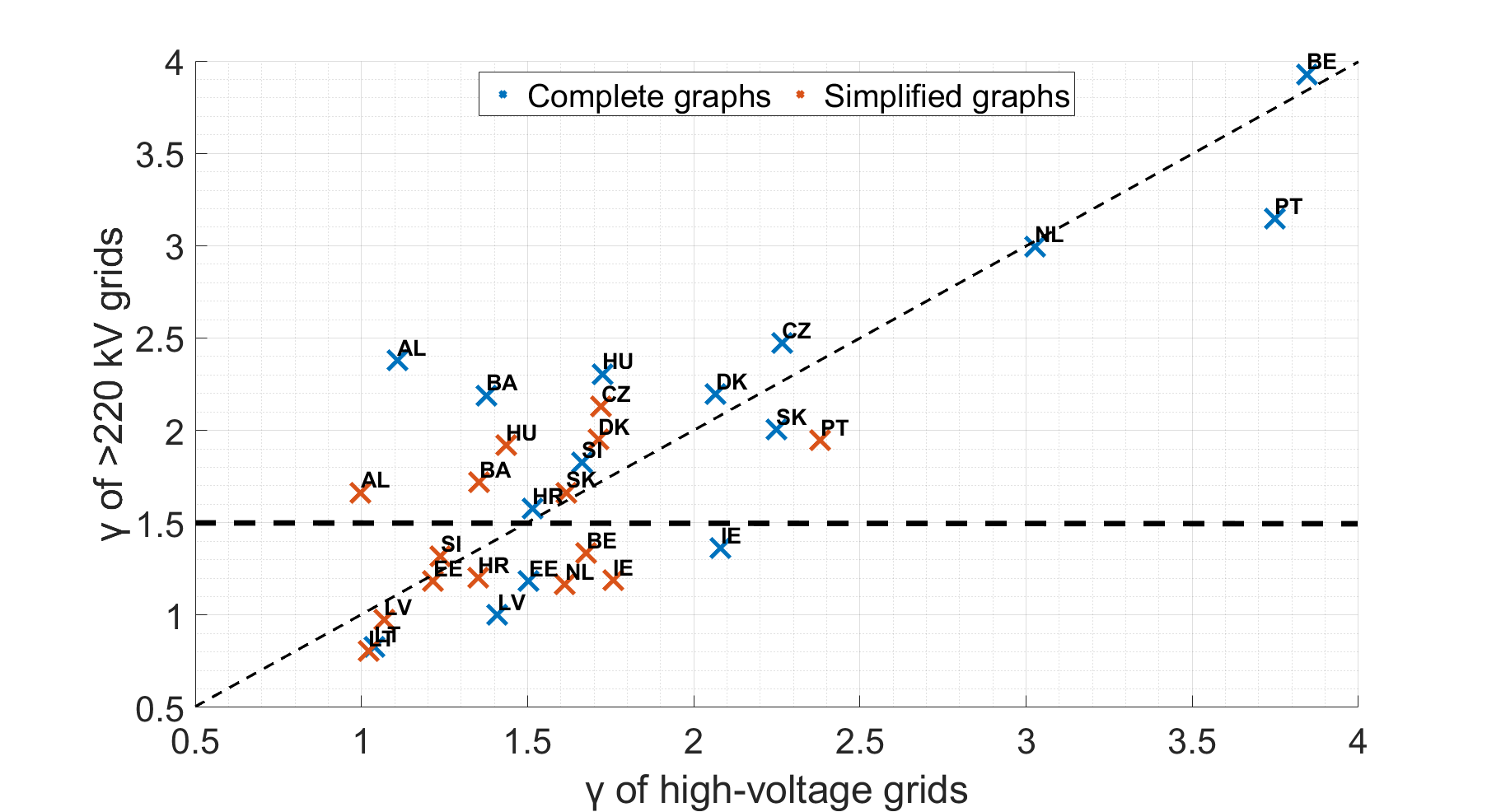}
     \caption{$\gamma$ values of exponential fits to the probability density function of node degree distributions. Blue indicates complete graphs, red indicates the simplified graphs. The $\gamma$ values of the literature are equivalent to the red markers compared to the y axis; 1.5 is a threshold for fragility as suggested by \cite{Sol2008}.}\label{fig:Gammas}
 \end{figure}

Figure~\ref{fig:Gammas} shows $\gamma$ values of the 15 examined networks. It can be seen that the inclusion of sub-transmission high-voltage parts of the network significantly modifies the node degree distributions, though this effect is not identical. In the case of 6 countries (EE, IE, LT, LV, PT, SK), $\gamma$ of the complete high-voltage grid is higher than the $\gamma$ of the transmission grid. The opposite is seen in 6 cases (AL, BA, CZ, DK, HU, SI), while only a minor difference is seen in the remaining 3 cases (BE, HR, NL).

\begin{figure}[H]
     \centering    \includegraphics[width=\columnwidth ]{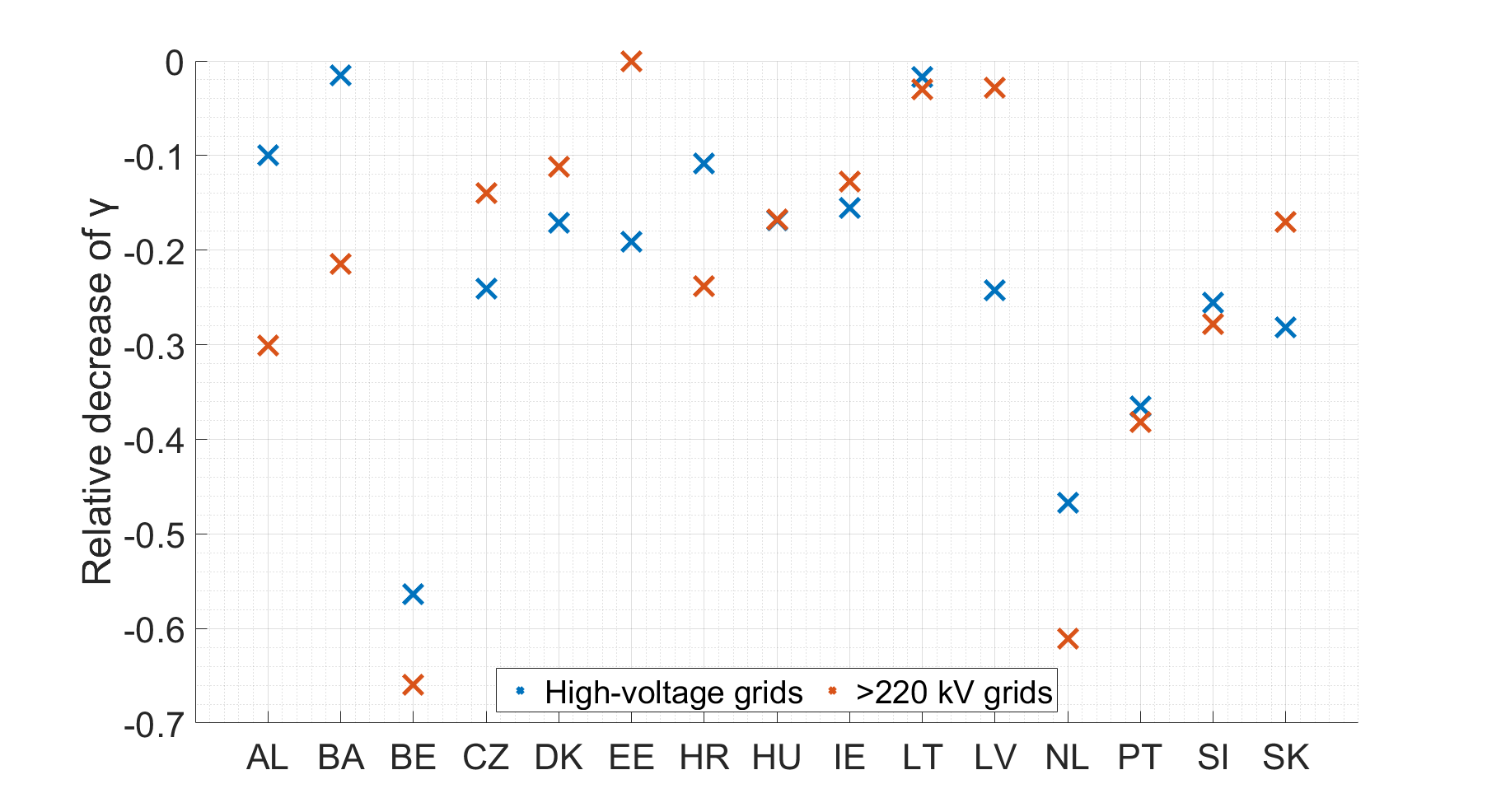}
     \caption{Relative decrease of $\gamma$ values due to simplification of the graph representations. Blue indicates topologies consisting of all voltage levels, red indicates transmission grid. Mean of the relative decrease are -0.2228, and -0.2306 respectively.}\label{fig:Gamma_decrease}
 \end{figure}

Simplifying the graphs decreases the values of $\gamma$, regardless of the voltage levels taken into consideration. This supports the claim that steeper probability density functions of node degree distribution are seen after redundant elements are removed. As Figure~\ref{fig:Gamma_decrease} shows, this relative decrease remains in the range below 30$\%$ for most of the networks. Significant outliers are the networks of BE, NL and PT, with the first two reaching values over 50$\%$; this is mostly the result of their well developed and strong high-voltage networks.

\begin{figure}[H]
     \centering    \includegraphics[width=\columnwidth ]{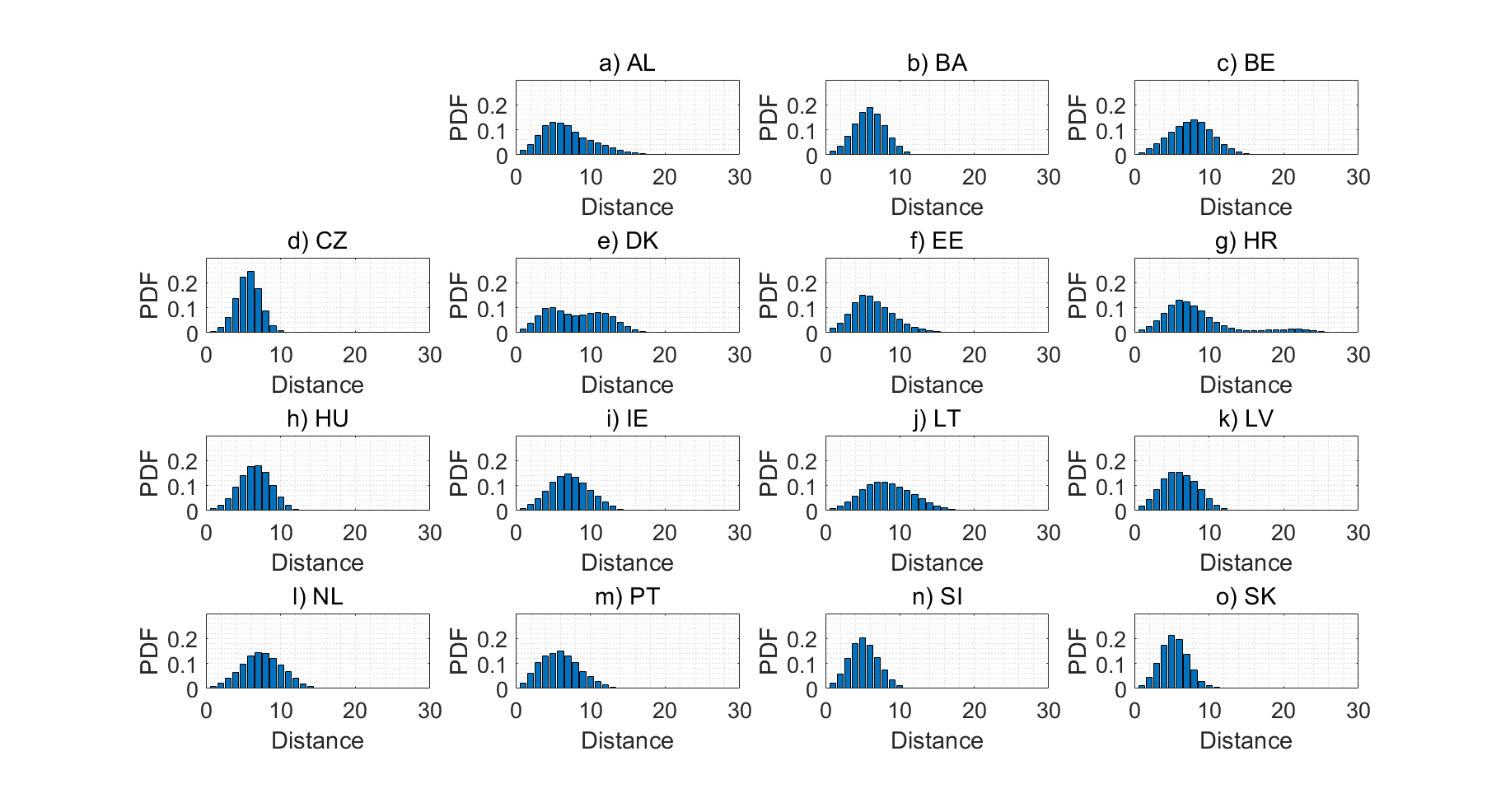}
     \caption{Probability density function of topological distance distribution of the networks. a) AL, b) BA, c) BE, d) CZ, e) DK, f), EE, g) HR, h) HU, i) IE, j) LT, k) LV, l) NL, m) PT, n) SI, o) SK}\label{fig:Distance_distribution}
 \end{figure}

Figure~\ref{fig:Distance_distribution} shows the topological distance distributions of the networks which are found to be normally distributed with most of them having a single peak indicative of one characteristic distance and uniform structure except for DK (Figure~\ref{fig:Distance_distribution}(e)) and HR (Figure~\ref{fig:Distance_distribution}(g)) which exhibit double peaks. This implies the presence of multiple clusters and substructures in these networks. In the case of DK this is the result of the Eastern and Western Denmark power systems connected through a single line (the Storebælt 400 kV HVDC link), while in the case of HR, the crescent-shaped geography of the country is the underlying reason.
Another property of the distributions that provide insight in the differences in the structures of the network is their variance i.e. most curves are narrower and some are wider (AL, BE, DK, EE, HR, LT). This indicates regularity (uniform distances between nodes) and diversity (variable distances and uneven structure) in topology.

\subsection{Effect of removals}\label{sec3.2}

\begin{figure}[H]
     \centering    \includegraphics[width=\columnwidth ]{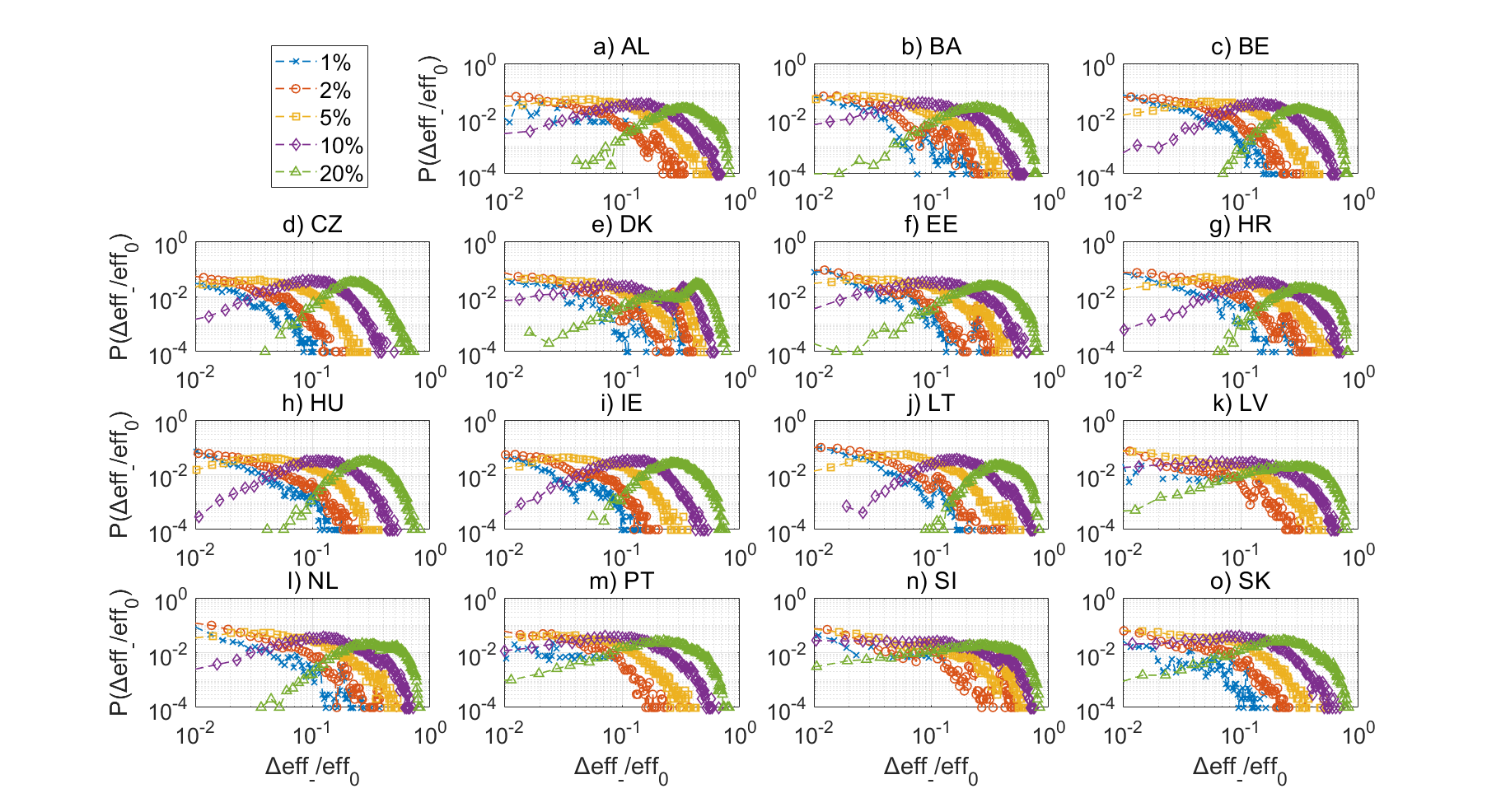}
     \caption{Probability density function of the decrease of network efficiency if given percentage of nodes are randomly removed. a) AL, b) BA, c) BE, d) CZ, e) DK, f), EE, g) HR, h) HU, i) IE, j) LT, k) LV, l) NL, m) PT, n) SI, o) SK}\label{fig:Node_removal_16pt}
 \end{figure}

\begin{figure}[H]
     \centering    \includegraphics[width=\columnwidth ]{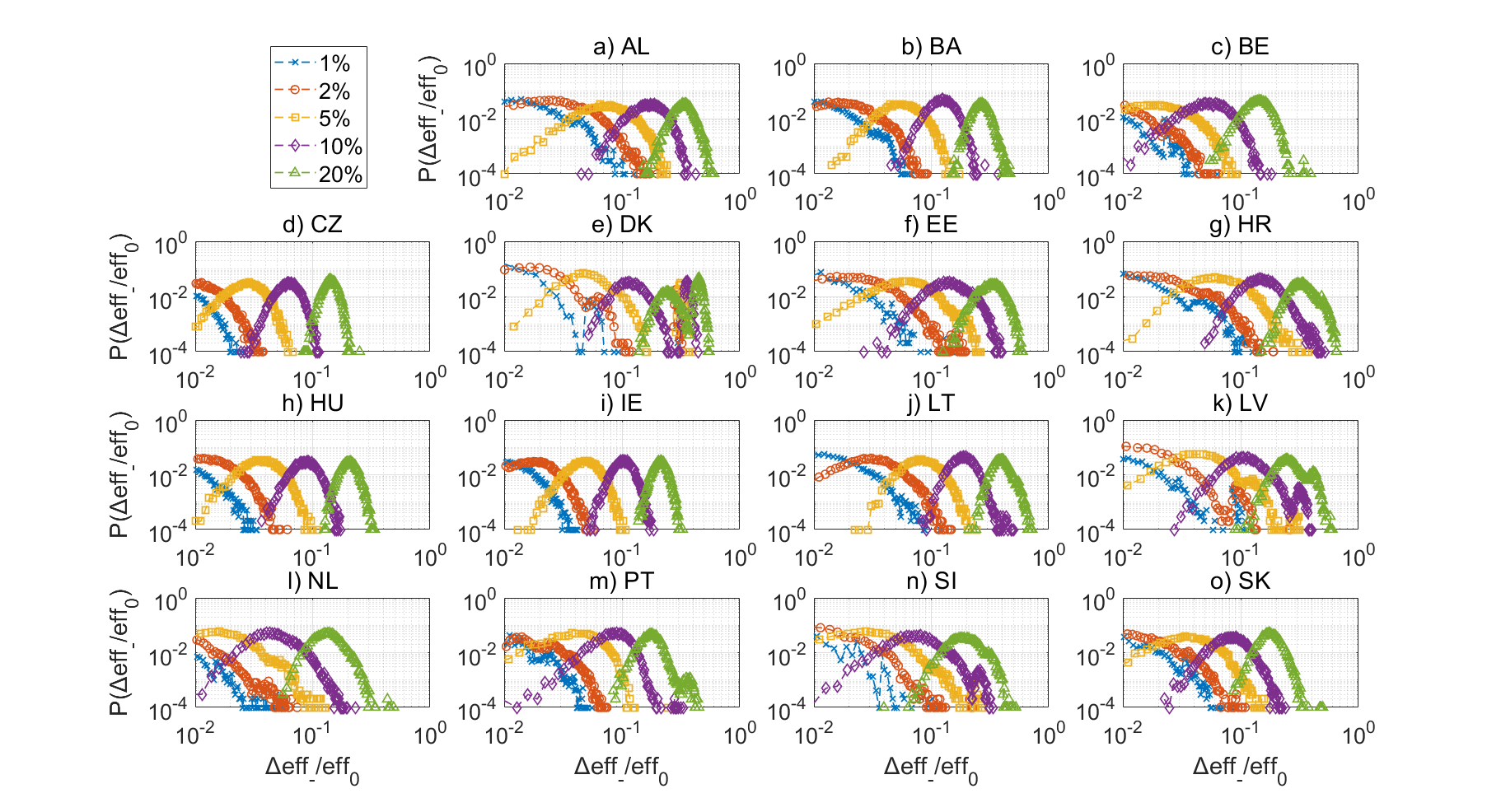}
     \caption{Probability density function of the decrease network efficiency if given percentage of edges are randomly removed. a) AL, b) BA, c) BE, d) CZ, e) DK, f), EE, g) HR, h) HU, i) IE, j) LT, k) LV, l) NL, m) PT, n) SI, o) SK}\label{fig:Edge_removal_16pt}
 \end{figure}

Figures \ref{fig:Node_removal_16pt} and \ref{fig:Edge_removal_16pt} show the probability density function of the decrease in network efficiency, when given percentage of nodes or edges are randomly removed from the network, respectively. All curves represent the event horizon of 10,000 realizations.

In Figure \ref{fig:Node_removal_16pt} when 1–2\% of the nodes are removed, we observe a higher probability of low efficiency decrease values, with the likelihood dropping slowly around the intermediate region and almost none at the high-efficiency region. This indicates that even a small fraction of node removal can target structurally important nodes (such as hubs or articulation points), sharply reducing network efficiency. At this level, the distributions can show multiple peaks, particularly for 1\%, reflecting presence of various critical elements that influence the network efficiency. Each peak corresponds to a recurring structural outcome depending on which nodes are removed: (i) if peripheral or low-degree nodes are removed, the network loses relatively few critical paths, and efficiency remains relatively high, producing a high-efficiency phase. Conversely, (ii) if a hub or a node connecting large subnetworks is removed, many shortest paths are disrupted, leading to a low-efficiency phase.

As more nodes are removed (5–20\%), expectedly, the decrease in the network efficiency will also progressively grow such that the distributions shift toward the low-intermediate efficiency region and persist there, until the high-efficiency region, where the probability drops rapidly. This plateau at the low-intermediate efficiency indicates that most realizations have already lost their global backbone connectivity, leaving only smaller connected components. The sharp drop at high efficiency highlights the fragility of the network under cumulative node loss. Notably, DK exhibits multiple peaks at all removal percentages, suggesting that its structure consistently produces distinct efficiency states regardless of how many nodes are removed.

In Figure~\ref{fig:Edge_removal_16pt} the distributions for the decrease in network efficiency is shown as some percentages of edges are removed randomly. Unlike node removal, where deleting even a few nodes can immediately remove entire hubs and all their connected edges thus fragmenting the network and producing highly variable efficiency outcomes, edge removal only deletes links, often leaving all nodes intact. This difference fundamentally changes how efficiency responds to perturbations. 

Upon the removal of 1\% of the network edges, the distributions show a high probability of low efficiency values, which is to be expected. Later on, this drops rapidly across intermediate and high-efficiency regions. Here, the distributions are often multi-peaked, reflecting heterogeneous outcomes. For example, some realizations lose critical shortcuts and drop to lower efficiency phases, while others preserve key paths and remain in higher-efficiency phases. By 2\%, the distributions tend to smooth out but still show a drop towards the intermediate efficiency values.

Of course, under higher percentage removals (5–20\%), the distributions become narrower with more distinct peaks (especially in the case of DK, HR, LV, PT, SI), often shifting toward high efficiency values. The narrow bell curve indicates that most network realizations settle around similar efficiency values, reflecting a relatively homogeneous response across trials. This happens because nodes remain intact, and alternative paths often compensate for the removal of edges, preventing significant efficiency drops. The appearance of distinct peaks at higher percentages of edge removal suggests that the network does not degrade uniformly, but instead settles into a set of stable efficiency states. Each peak corresponds to a recurring structural outcome that arises across many random realizations. For example, in some cases, edge removals may primarily affect redundant or peripheral links, leaving the core shortcuts intact, producing a phase of higher efficiency. In other cases, removals may target a handful of longer-range or higher-betweenness edges, forcing paths to reroute through slightly longer routes, resulting in a phase of moderately lower efficiency. This time, we can observe DK and LV exhibiting multiple distinct peaks for all percentages of edge removals, suggesting several stable stages at all levels of node loss, driven by its structural heterogeneity, hub distribution, and connectivity patterns.

Despite the discussed variations, the decrease in efficiency values associated with these stable states remain relatively high compared with the node-removal scenario. This resilience arises because the nodes themselves are preserved, ensuring that the overall connectivity is not fragmented. Even with substantial edge loss, most pairs of nodes remain reachable, and alternative paths can compensate for missing links. Thus, while edge removal produces multiple efficiency modes reflecting structural heterogeneity, the network continues to function in a globally efficient manner compared with the more severe disruptions caused by node removal.

\subsection{Performance}\label{sec3.3}

\begin{figure}[H]
     \centering    \includegraphics[width=\columnwidth ]{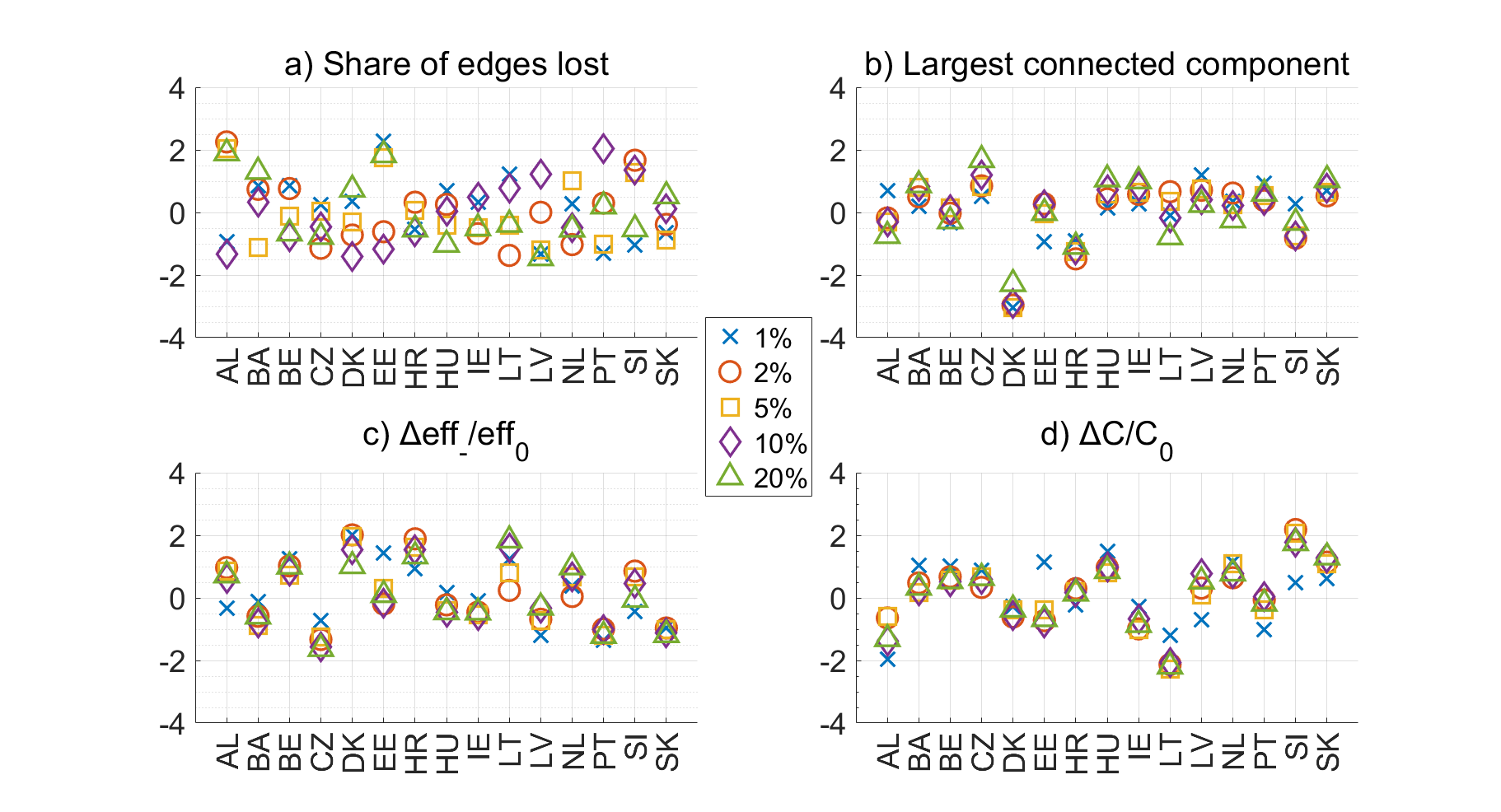}
     \caption{Standardized performance of the networks across the four metrics if given percentage of nodes are randomly removed}\label{fig:Node_removal_metrics}
 \end{figure}

\begin{figure}[H]
     \centering    \includegraphics[width=\columnwidth ]{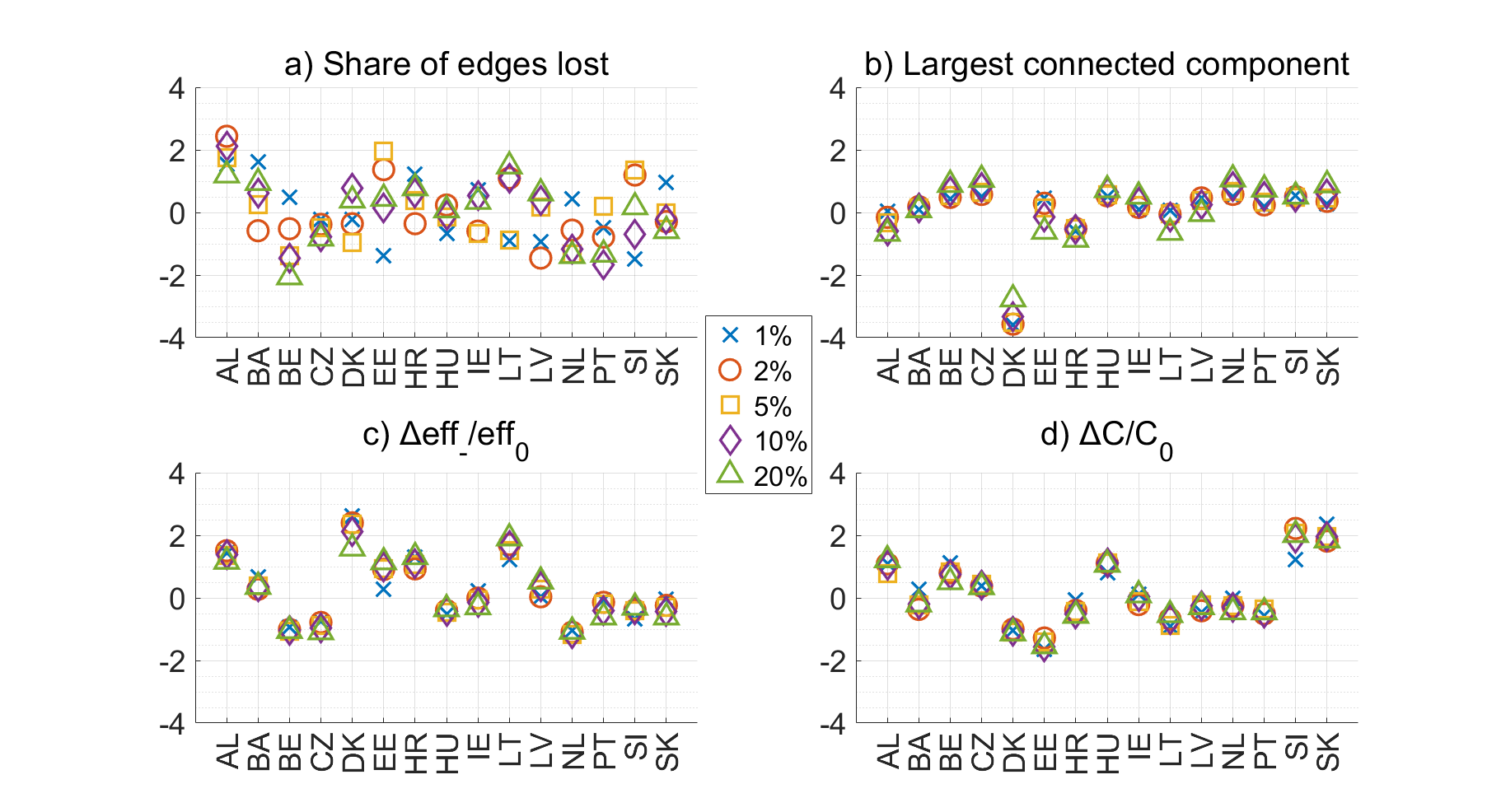}
     \caption{Standardized performance of the networks across the four metrics if given percentage of edges are randomly}\label{fig:Edge_removal_metrics}
 \end{figure}

Figures~\ref{fig:Node_removal_metrics} and \ref{fig:Edge_removal_metrics} shows the performance of the networks based on various metrics (a) share of edges lost (b) largest connected component LCC (c) change in network efficiency $\Delta eff_{-}/eff_0$ (c) change in clustering coefficient $\Delta C/C_{0}$ upon the removal of nodes and edges.
The share of edge lost metric hints at the extent of damage in the network when it experiences disturbances. Here, a network is considered to perform well if the share of edges lost is not high. On the other hand, the size of the largest connected component (LCC) is the number of nodes in the connected component after the disturbances. If the LCC shrinks significantly after the failures, the network is fragile and less resilient. The change in efficiency with respect to the baseline efficiency $\Delta eff_{-}/eff_0$ captures how flow changes after the perturbations, which tells us about the network's adaptive resilience. Finally, the change in clustering $\Delta C/C_{0}$ reflects local redundancy and robustness such that the change here tells us about structural vulnerability.

When networks face random shocks through node removal, their ability to hold together depends heavily on which aspect of the network we look at. These findings highlight the different ways in which networks absorb or fail to absorb shocks, depending on the structural property under examination. Looking first at the share of edges lost (shown in Figure~\ref{fig:Node_removal_metrics}(a)). Here, the higher they are positioned vertically, the worse their performance. For example, country-level patterns reinforce this interpretation: LV consistently outperformed others (at 1\%, 5\%, and 20\%), suggesting structural features that enable it to retain edges even under high levels of stress, whereas EE and AL consistently underperformed, highlighting their vulnerability to edge loss.

In Figure~\ref{fig:Node_removal_metrics}(b)), the LCC revealed a different dimension of resilience. This time, the higher they are positioned vertically, the better the performance. Up to 10\% node removal, most countries’ networks showed remarkable cohesion, with fragmentation largely absent. This suggests that global connectivity is relatively stable under modest shocks. However, beyond this threshold, performance collapsed: by 20\%, fewer than half of the networks remained cohesive. Notably, CZ consistently maintained large connected components across multiple removal levels, pointing to a robust capacity for preserving global structure. In contrast, DK consistently fragmented quickly, positioning it as a clear outlier. Together, these results imply that while global connectivity is relatively resistant to initial shocks, there exists a tipping point beyond which collapse is sudden and severe.

We also look into the decrease in the efficiency $\Delta eff_{-}/eff_0$ as more nodes are removed in the network. Figure~\ref{fig:Node_removal_metrics}(c) highlights clear differences in country performance. Here, the higher they are positioned vertically, the worse their performance. At the 1\% minimum threshold, PT emerges as the best performer, showing the smallest efficiency drop, while DK performs the worst. As the minimum threshold increases to 2\%, 5\%, 10\%, and 20\%, the CZ consistently demonstrates resilience, maintaining its position as the best performer with the lowest efficiency decrease. In contrast, DK continues to underperform at the 2\% and 5\% thresholds, but at higher levels (10\% and 20\%), LT replaces DK as the worst performer, showing the largest efficiency losses. This pattern indicates that while the CZ maintains stability under increasing thresholds, DK and LT struggle significantly, with LT’s efficiency being particularly vulnerable at higher thresholds.

Finally, Figure~\ref{fig:Node_removal_metrics}(d) shows the performance of the countries across all percentages in terms of the relative change in clustering coefficient. The results of node removal and its effect on clustering reveal consistent trends in network resilience. Similarly, for this metric the higher they are vertically, the worse the performance. At the 1\% minimum removal threshold, AL performs best, experiencing the smallest change in clustering, while HU is the most negatively affected. However, from the 2\% threshold onward, LT emerges as the best performer, maintaining minimal changes in clustering as removal levels increase. On the other hand, SI consistently performs worst from the 2\% through the 20\% thresholds, showing the greatest clustering disruptions. This suggests that while AL demonstrates strong resilience at the very lowest level of node removal, LT is more robust under sustained or higher removal conditions. In contrast, SI’s clustering is highly vulnerable even to modest node removal, and HU shows weakness at the very beginning.

These results emphasize the multi-dimensional nature of network resilience. Depending on the metric, the same network can appear robust, fragile, or adaptive. Edges and LCC reveal thresholds where collapse accelerates, efficiency shows immediate and universal sensitivity, and clustering exposes local-level vulnerabilities masked by global stability. These findings suggest that assessing resilience requires a holistic perspective: focusing on any single metric risks mischaracterizing the true vulnerabilities of a system.
 
Figure~\ref{fig:Edge_removal_metrics} provides important insights into how networks respond to the direct removal of edges, as compared to node removal. The first striking observation comes from the share of edges lost (Figure~\ref{fig:Edge_removal_metrics}(a)). BE stood out as a recurring best performer (at 5\% and 20\%), maintaining resilience even when edge density dropped. Conversely, AL showed persistent vulnerability, ranking among the worst at both 2\% and 10\%, while EE underperformed again this time in the edge removal scenario, particularly at low levels of removal (5\%).

The LCC told a more robust story in Figure~\ref{fig:Edge_removal_metrics}(b). While performance gradually declined as removals increased, from 87\% of networks performing well at 1\% removal down to 60\% at 20\%, most countries nevertheless retained large connected components even under high stress. This result contrasts with the share of edges lost metric, where fragility was far more evident at 20\%. In other words, networks tend to preserve their global structure longer than they preserve their edge density. Among individual countries, the NL demonstrated exceptional robustness, dominating across 2\% and above. SI only led at very low removal levels (1\%), while DK consistently performed worst, fragmenting even under minimal shocks, similar to its behavior under node removal. DK thus remains a clear outlier in its susceptibility to fragmentation.

The impact of edge removal on the efficiency (Figure~\ref{fig:Edge_removal_metrics}(c)) shows a strong divide between the most resilient and most vulnerable countries. Similar to the node removal scenario, the higher a country is positioned vertically, the worse is its performance. For example, at the 1\%, 2\%, 5\%, and 10\% thresholds, the NL consistently stands out as the best performer, experiencing the smallest efficiency decrease, while DK is the weakest, suffering the largest losses across these levels. However, at the 20\% threshold, the pattern shifts: the CZ emerges as the most resilient, while LT replaces DK as the worst performer, showing the steepest efficiency decline. This progression suggests that the NL is highly robust under lower to moderate levels of edge removal, but the CZ outperforms at more extreme levels. In contrast, DK is particularly vulnerable at lower thresholds, whereas LT’s efficiency becomes especially fragile under severe edge loss. Note that CZ and LT were also the best and worst performers in the node removal scenario at 20\%.

Clustering dynamics revealed additional nuance. Unlike in the node removal case, where clustering sometimes improved under small shocks, edge removal tended to produce more deterioration as can be seen in Figure~\ref{fig:Edge_removal_metrics}(d). At the outset, both node and edge removal scenarios saw around half of the networks performing poorly, but as removals increased, the edge removal case drove a larger decline in clustering resilience. This suggests that edges play a disproportionately important role in maintaining local structure: their loss directly undermines tightly knit subgroups.  The effects of edge removal on clustering reveal a consistent pattern of resilience and vulnerability among countries. Across all thresholds, EE consistently emerges as the best performer, showing the smallest change in clustering and indicating strong robustness to edge removal. The worst performers, however, vary depending on the threshold. At 1\% and 10\%, SK experiences the greatest disruption, while at 2\%, 5\%, and 20\%, SI becomes the most negatively affected. This suggests that EE’s clustering structure is remarkably stable regardless of the severity of edge removal, whereas both SK and SI display significant weaknesses. SK is especially vulnerable at lower and moderate levels, while SI’s clustering deteriorates more consistently as edge removal intensifies.

\begin{figure}[H]
     \centering    \includegraphics[width=\columnwidth ]{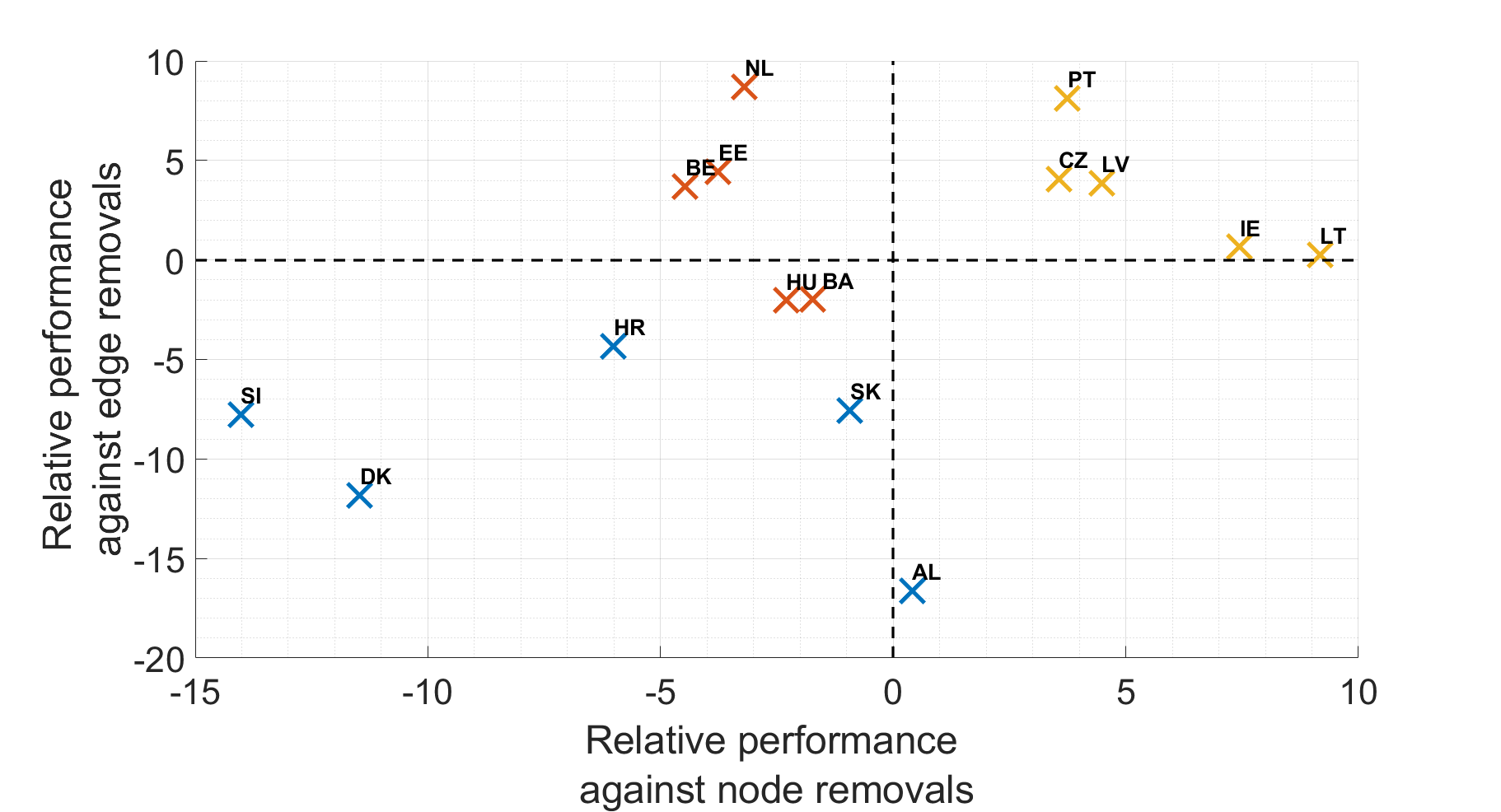}
     \caption{Performance of the networks against node and edge removals. Positive values indicate better performance.}\label{fig:Performance_clusters}
 \end{figure}

The results, so far,  reveal two fundamentally different resilience regimes. Node removal exposes the acute fragility of networks: even small shocks can trigger sharp efficiency collapses, with outcomes highly dependent on whether structurally important hubs are lost. This produces heterogeneous, multi-peaked efficiency states that reflect deep structural vulnerability. Edge removal, by contrast, preserves the nodes and therefore enables adaptation. While small shocks still produce instability, networks quickly stabilize into a narrower set of efficiency states, often retaining higher functional performance even under heavy losses. Thus, node removal highlights catastrophic vulnerability driven by structural anchors, while edge removal illustrates adaptive resilience supported by alternative routing. Taken together, these contrasting outcomes suggest a design principle for resilient systems: safeguarding critical nodes is paramount to avoid collapse, while redundant or distributed edge structures can buffer against connectivity loss and sustain functionality. Having said this, it therefore makes sense to investigate how the countries fare in both types of failures. In Figure~\ref{fig:Performance_clusters}, we show a composite plot of the performance of the countries under node and edge removals.

Countries that did poorly under node removal scenarios (left plane) have power networks that are overly reliant on hubs or key nodes making them structurally vulnerable. On the other hand, countries that performed poorly under edge removal scenarios (lower plane) indicate a reliance on certain power lines that are considered inadaptive. Therefore, countries that perform well under both scenarios, which suggest balanced, redundant, and resilient structures, can be found in the upper-right quadrant of Figure~\ref{fig:Performance_clusters}; those that behave otherwise are found in the lower-left quadrant.

In Figure~\ref{fig:Performance_clusters}, the datasets are categorized into three groups. Group I (blue) includes the worst-performing power networks, characterized as structurally vulnerable and inadaptive (SI, DK, SK, AL, HR). These countries repeatedly underperform across several metrics, revealing both global and local fragility. SI exemplifies this duality: although it briefly performs well in LCC at 1\% edge removal, it consistently ranks worst in clustering change ($\Delta C/C_{0}$) under both node (2–20\%) and edge (2, 5, 20\%) removals. This suggests that while SI’s global structure can momentarily hold together, its local redundancy collapses quickly, pointing to an over-reliance on a few key connections. DK is the most consistently fragile: it ranks worst in LCC across all thresholds, and in efficiency at low removal levels (1–5\% node removal, 1–10\% edge removal). This pattern is directly linked to the Eastern and Western Denmark power systems being connected through a single line (the Storebælt 400 kV HVDC link). SK also exhibits weakness in clustering (worst at 1\% and 10\% edge removal), reflecting poor local robustness and limited backup connectivity. On the other hand, AL is grouped here as it performed worst in edge loss at several thresholds (2, 5\% node removal; 2, 10\% edge removal). However, it did perform well in terms of clustering at 1\% node removal, suggesting strong local resilience. However, AL does not sustain this advantage in global measures such as LCC or efficiency, indicating that while its networks can absorb small shocks locally, they lack global backbone stability. HR, similarly, exhibits structural weaknesses that prevent it from joining the more resilient groups. Overall, Group I networks are marked by fragility, hub dependence, and low redundancy, which make them highly vulnerable to both random failures and targeted disruptions.

Group II (orange) represents networks with intermediate performance (NL, EE, BE, HU, BA). NL performs best in efficiency changes (1–10\%) and LCC at 2–20\%  under edge removal. This indicates that NL can retain connectivity even after large (at 20\%) edge perturbations, but it cannot sustain its performance in terms of efficiency under this level of attack. EE demonstrates a duality: it excels in clustering ($\Delta C/C_{0}$, 1–20\% edge removal) but performs worst in edge loss at 1\% and 20\% node removal and 5\% edge removal, showing robust local redundancy and the ability to re-route flows, but is not globally robust; thus, it does not dominate efficiency or LCC measures. BE performs well in edge loss (best at 5\% and 20\% edge removal), reflecting resilience to link failures, likely due to the availability of alternative connections. HU performs worst in clustering at 1\% node removal, suggesting local fragility even at minimal shocks. BA, meanwhile, achieves the worst performance in edge loss at 1\% edge removal, which suggests initial weakness and the lack of stability at higher thresholds where its structure cannot sustain performance. Taken together, Group II networks highlight trade-offs: some preserve efficiency at the expense of cohesion, others protect local redundancy while losing global adaptability.

Group III (yellow), in the upper-right quadrant, includes adaptive and structurally resilient networks (PT, CZ, LV, LT). These maintain both global cohesion and local redundancy under stress, though with variations in how resilience is achieved. PT illustrates the balance of strengths and weaknesses: it performs best in efficiency at 1\% node removal and edge loss at 10\% edge removal, but is worst in edge loss at 10\% node removal, showing that while its structure is strong under edge failures, it is more fragile under node failures. CZ is the most resilient overall: consistently best in LCC (2–20\% node removal) and efficiency (2–20\% node removal, 20\% edge removal), it reflects a highly redundant and well-connected backbone that allows the network to preserve both cohesion and adaptability under stress. On the other hand, LV is best in LCC at 1\% node removal and in edge loss at multiple thresholds (1, 5, 20\% node removal; 2\% edge removal). This suggests that its core holds together as failures accumulate. LT demonstrates a dual profile: it performs best in clustering under node removal (2–20\%) and edge loss (2\% node removal), suggesting strong local redundancy, but is worst in efficiency (10–20\% node removal; 20\% edge removal) and edge loss (20\% edge removal). This reflects a structure that maintains local cohesion but loses global adaptability under stress, with efficiency dropping steeply as connectivity deteriorates. Group III networks thus show the capacity to absorb shocks, either through strong redundancy (CZ, EE) or adaptability in efficiency (PT), though some (LT) remain vulnerable in specific dimensions.

\section{Discussion}
\label{sec:sec4}

Overall, the comparison between node and edge removal scenarios highlights important asymmetries. Edge removal produces higher fragility in clustering but less collapse in global connectivity, since nodes remain present to anchor the network. The persistence of country-level patterns, such as DK’s fragmentation under both removal scenarios, NL and CZ's efficiency resilience; LT and EE’s clustering strength, suggests that some vulnerabilities and strengths are deeply embedded in the underlying structure of these networks. Taken together, these findings reinforce the need for multi-metric assessments: focusing solely on connectivity or edge retention would obscure the hidden vulnerabilities in efficiency and clustering that emerge even under minimal shocks.

The probability distributions of the decrease in efficiency help explain the country-level performance outcomes observed under node and edge removal. In the node-removal scenario (Figure~\ref{fig:Node_removal_16pt}), the distributions show that even removing 1–2\% of nodes leads to a high likelihood of low efficiency values, with multiple peaks appearing due to structural heterogeneity. This variability reflects the fact that removing hubs or articulation points can sharply fragment connectivity and drop efficiency, whereas removing peripheral nodes has only minor effects. The multiple peaks observed, particularly for DK, are consistent with its poor performance metrics: DK frequently emerges as one of the worst performers in efficiency and LCC resilience, showing that its structure is susceptible to which nodes are lost. Conversely, countries such as PT (at 1\%) and CZ (from 2\% onwards) appear at the favorable ends of the efficiency performance metrics.

In contrast, distributions under the edge-removal scenario (Figure~\ref{fig:Edge_removal_16pt}) highlight the different structural dynamics at play. Since edge removal preserves nodes, connectivity is rarely lost entirely, and efficiency distributions shift more gradually. At 1\%, the distributions still show high probabilities of low efficiency values and multi-peaked structures, but as removal increases (5–20\%), the distributions narrow and form distinct peaks. This stabilization explains why countries such as NL consistently emerge as the best performers at low and moderate levels: its structure maintains high efficiency by leveraging redundant connections, even when shortcuts are lost. At higher removal levels (20\%), CZ overtakes NL, reflecting a shift in the PDF distribution. Similarly, LT appears as the worst performer under severe edge loss, consistent with the probability distributions showing how certain networks settle into much lower efficiency states. The clustering results also connect directly: EE’s stability across all thresholds reflects consistently narrow distributions, while SI’s and SK’s vulnerabilities correspond to cases where efficiency and clustering are most disrupted by missing edges.

The distributions of the decrease in efficiency illuminate why some countries repeatedly perform well or poorly across the resilience metrics. Node removal produces highly variable outcomes due to the structural importance of hubs, leading to multiple efficiency phases and sudden collapses, which explains the sharp performance contrasts between robust networks like CZ and fragile ones like DK. Edge removal, by contrast, generates multiple peaks at higher thresholds for some countries, indicating higher resilience since connectivity is preserved. This explains why networks such as NL and EE dominate performance rankings, while structurally weaker systems like SI and SK repeatedly underperform.

 \begin{figure}[H]
     \centering    \includegraphics[width=\columnwidth ]{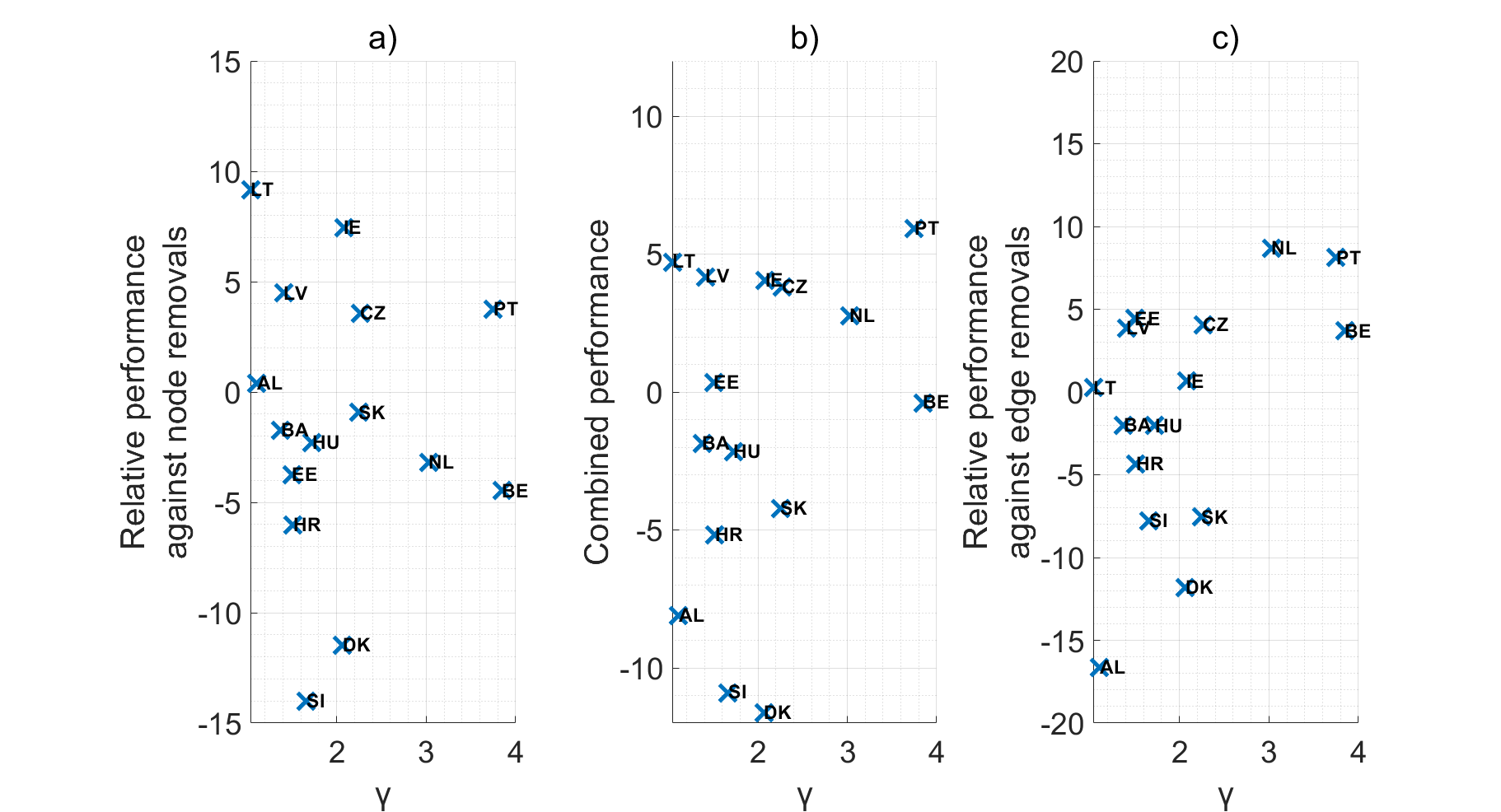}
     \caption{Relative performance of the networks against various removals, plotted against the $\gamma$-s of their node degree distribution}\label{fig:Lambdas_vs_performance.png}
 \end{figure}

Figure~\ref{fig:Lambdas_vs_performance.png} a comparative analyses of the datasets showing the relationship of their $\gamma$ values to their performances under both node and edge removal scenarios. It is well-known that for power grids the degree distributions behaved exponentially. Under intentional attacks, networks that have high-degree nodes ($\gamma > 1.5$) are considered fragile~\cite{Sol2008}; and robust otherwise. This can be expected, as in sparse networks such as the power grids, the absence of hubs makes intentional attacks ineffective.

From \ref{sec:sec3} of this present work, we saw that the inclusion of more power lines in the networks had varying effects to the $\gamma$ values of the networks. To reiterate, the complete graphs of EE, IE, LT, LV, PT, and SK had $\gamma_{HV} > \gamma_{ < 220kV}$. Meanwhile, the opposite is observed for AL, BA, CZ, DK, SI.

Figure~\ref{fig:Lambdas_vs_performance.png}(a) and (c) show the results under random node and edge removal strategies, respectively. Under random attacks, networks with higher $\gamma$ are more robust because of the multiple links i.e. removing a few nodes rarely fragments the network. This explains why most of the countries that belong to Group III (Figure~\ref{fig:Performance_clusters}) and were observed to have $\gamma_{HV} > \gamma_{ < 220kV}$ performed better under random node removal as can be seen more apparently in Figure~\ref{fig:Lambdas_vs_performance.png}(a). However, there are cases where we found sparse networks ($\gamma < 1.5$) that performed well even under random attacks on nodes (AL, LT, and LV). This may be due to the fact that most nodes have low degree and there are only a few hubs that are rarely hit under random removals. Additionally, because there are only a few connections per node, the average path length may be longer but the connectivity is not easily destroyed.

In the case of random edge removal (Figure~\ref{fig:Lambdas_vs_performance.png}(c), this time we see countries that performed well and they are identified to fall under Group II (NL, BE, EE) and Group III (IE, LT, LV, PT) and were observed to either have $\gamma_{HV} \approx \gamma_{ < 220kV}$ or $\gamma_{HV} > \gamma_{ < 220kV}$, respectively. This time, EE, LV, and LT have $\gamma < 1.5$ indicating that they are sparse. Their unexpected good performance under random edge removal may be due to the networks being homogeneous and the fact that the removal may be spread out which means that the nodes are rarely disconnected simultaneously. The same trend is observed for the combined performance shown in (Figure~\ref{fig:Lambdas_vs_performance.png}(b). Meanwhile, the countries from Group I, while they can be considered to have a moderate $\gamma$-values (between $\approx 1-2.5$), consistently performed poorly against both attack strategies.

\section{Conclusion}
\label{sec:sec5}

This study demonstrated that the vulnerability of European high-voltage grids should be evaluated using composite performance metrics instead of single ones. Node and edge removals exposed distinct structural weaknesses: the loss of nodes tends to fragment the networks abruptly through the removal of hubs, while the loss of edges degrades efficiency more gradually while still undermining clustering. The persisting examples of country-specific patterns (e.g. the efficiency of the Netherlands' and Czechia's grid, the clustering strength of Estonia and Lithuania) show that vulnerabilities are embedded in the fabric of the network.

The relation between the decay rates of node degree distributions and performance metrics further strengthen these findings. Systems with higher decay constants typically resist random failures, while more sparse networks can still perform well under certain conditions. These latter outcomes are results of the homogeneity of these networks, with lower number of critical hubs. We also observed that networks with a mid-range decay tend to underperform in multiple scenarios, suggesting that fragility is not a simple monotonic function.

Our cross-country analysis highlights the importance of multi-metric assessments that integrate degree distributions, efficiency, clustering, and robustness tests. Considering these dimensions of grid vulnerability together, we can get a more reliable and clear picture on how well certain networks handle failures.

\section{Acknowledgement}
\label{sec:acknowledgements}
Bálint Hartmann acknowledges the support of the Bolyai János Research Scholarship of the Hungarian Academy of Sciences (BO/131/23), and the National Research Excellence Programme of the National Research, Development and Innovation Office (ADVANCED 150405).

% \appendix
%\section{Appendix A} \label{app1}

\bibliographystyle{elsarticle-num}
\bibliography{references.bib}
\end{document}